\documentclass{article}

\usepackage[final,nonatbib]{nips_2017}
\usepackage[numbers]{natbib}

\usepackage[utf8]{inputenc} 
\usepackage[T1]{fontenc}    
\usepackage{url}            
\usepackage{booktabs}       
\usepackage{amsfonts}       
\usepackage{nicefrac}       
\usepackage{microtype}      
\usepackage{graphicx,float}
\usepackage{amssymb,amsmath,amsthm}
\usepackage{cleveref}
\usepackage{multirow,sparklines}
\usepackage{soul}
\usepackage{hyperref}       
\newcommand{\ignore}[1]{}

\title{Convolutional neural networks for structured omics: OmicsCNN and the OmicsConv layer}
\author{
\begin{tabular}{c}
G. Jurman\textsuperscript{1,*}, V. Maggio\textsuperscript{1}, D. Fioravanti\textsuperscript{2}, Y. Giarratano\textsuperscript{3}, I. Landi\textsuperscript{1,4}\\
M. Francescatto\textsuperscript{1}, C. Agostinelli\textsuperscript{4}, M. Chierici\textsuperscript{1}, M. De Domenico\textsuperscript{1,5}, C. Furlanello\textsuperscript{1}
\end{tabular}\\
\begin{tabular}{ll}
\textsuperscript{1} Fondazione Bruno Kessler, Trento (IT)  & \textsuperscript{2} MPI Intelligent Systems, Tuebingen (DE)\\
\textsuperscript{3} University of Edinburgh (UK) & \textsuperscript{4} University of Trento (IT)\\
\textsuperscript{5} Universitat Rovira e Virgilii, Tarragona (ES)\\
\textsuperscript{*} Corresponding author: \texttt{jurman@fbk.eu}
\end{tabular}
}

\begin{document}
\maketitle

\begin{abstract}
Convolutional Neural Networks (CNNs) are a popular deep learning architecture widely applied in different domains, in particular in classifying over images, for which the concept of convolution with a filter comes naturally. 
Unfortunately, the requirement of a distance (or, at least, of a neighbourhood function) in the input feature space has so far prevented its direct use on data types such as omics data. 
However, a number of omics data are metrizable, \textit{i.e.}, they can be endowed with a metric structure, enabling to adopt a convolutional based deep learning framework, \textit{e.g.}, for prediction. 
We propose a generalized solution for CNNs on omics data, implemented through a dedicated Keras layer. 
In particular, for metagenomics data, a metric can be derived from the patristic distance on the phylogenetic tree. 
For transcriptomics data, we combine Gene Ontology semantic similarity and gene co-expression to define a distance; the function is defined through a multilayer network where 3 layers are defined by the GO mutual semantic similarity while the fourth one by gene co-expression. 
As a general tool, feature distance on omics data is enabled by OmicsConv, a novel Keras layer, obtaining OmicsCNN, a dedicated deep learning framework. 
Here we demonstrate OmicsCNN on gut microbiota sequencing data, for Inflammatory Bowel Disease (IBD) 16S data, first on synthetic data and then a metagenomics collection of gut microbiota of 222 IBD patients. 
\end{abstract}

\section{Introduction}
\label{sec:introduction}
The diversity and the complex structure of biological data represent both a challenge and an opportunity Deep Learning (DL) techniques~\cite{ching17opportunities}.
In fact, the superiority of deep neural network approaches is demonstrated on a growing number of biological and clinical tasks~\cite{mamoshina16applications,chaudhary17deep,zacharaki17prediction}.
Nonetheless, the DL potential is clearly being limited on omics data: open issues are transversal to other domains, \textit{e.g.}, limited availability of data, interpretability, hyperparameters tuning~\cite{min16deep}, but overall there is a need for architectures designed to use biological knowledge and underlying structure. 
As a consequence, the direct application of deep networks to microarray, genetic or other omics can be disappointing, as found by Ditzler and coauthors~\cite{ditzler15multilayer} that claimed that deep learning ''may not be suitable for metagenomic application''. 
 
In order to bridge this gap, we propose a short path to the application of Convolutional Neural Network (CNNs) architectures to omics data. CNNs is the elective solution for image classification~\cite{lecun98gradient,krizhevsky12imagenet}, based on convolving subsets of the input image  with different filters, where the key concept is the definition of a set of nearest neighbours for a given pixel. 
Using the same architecture for non-image data requires the availability of an analogous proximity measure between features.
 
As a first step, we show how to embed omics data in a metric space: in particular, we consider the case of metagenomics and transcriptomics. 
In the former case, the metric is a tree distance on the phylogenetic tree, while in the latter case we propose a more complex structure based on the diffusion distance on a 4 layers multilayer network~\cite{dedomenico17diffusion}. 
Given the distance, the neighbour detection comes naturally.
The convolutional layer combined with the neighbours detection algorithm is operatively implemented as a novel Keras layer~\cite{chollet15keras} called OmicsConv. 
The overall architecture OmicsCNN then consists of a stack of OmicsConv layers first flattened then terminating with a Fully Connected (Dense) and a final classification layer. 
The environment is completed by an experimental setup for unbiased prediction, \textit{i.e.}, a 10$\times$5-fold Data Analysis Protocol (DAP) developed for unbiased biomarker identification within the US-FDA led initiatives MAQC/SEQC~\cite{maqc10maqc,seqc14comprehensive}. 
 
As a case study, we conclude with a comparison of the OmicsCNN DAP on the metagenomics IBD dataset~\cite{sokol17fungal} with other classical and DL algorithm, where the OmicsCNN is applied in a domain adaptation strategy, freezing the first layers after they have been trained on a large synthetic dataset.
Further details on the metagenomics case are described in~\cite{fioravanti17phylogenetic}.
 
\begin{figure}[!t]
\centering\includegraphics[width=0.9\textwidth]{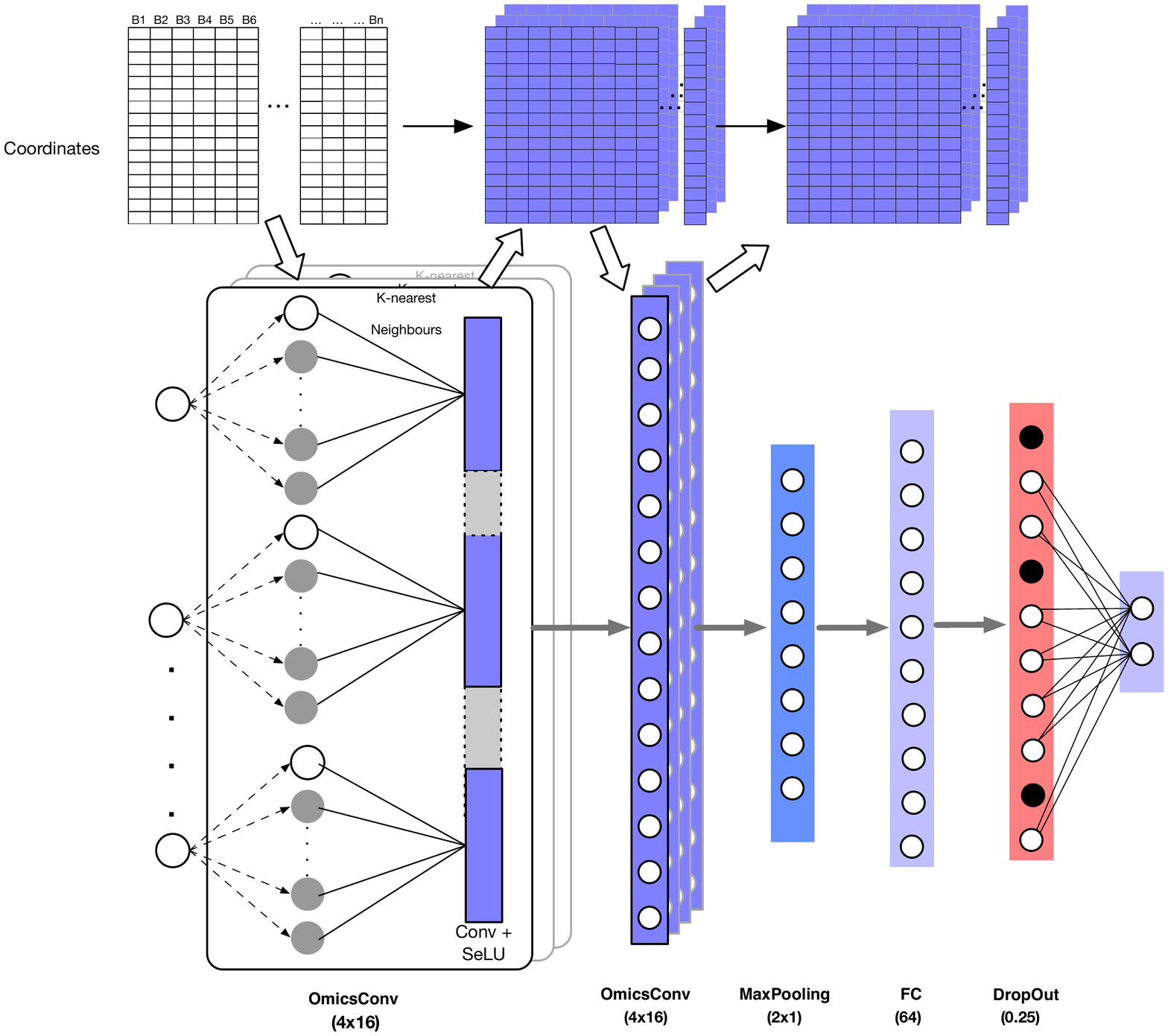}
\caption{The structure of OmicsCNN. In this configuration, OmicsCNN is composed by two OmicsConv layers followed by a Fully Connected layer before decision.}
\label{fig:omicscnn}
\end{figure}

\section{Metrizing omics data}
\label{sec:metrizing}
Differently from images, a natural distance between features is not immediately defined for most of omics data types. 
Although an indirect transformation into images of outcomes from sequencing has been proposed, we seek here to start a principled development of methods that can apply to some of the most interesting omics data types for clinical application and scientific discovery.
 
\begin{figure}[!t]
\centering
\begin{tabular}{cc}
\includegraphics[width=0.5\textwidth]{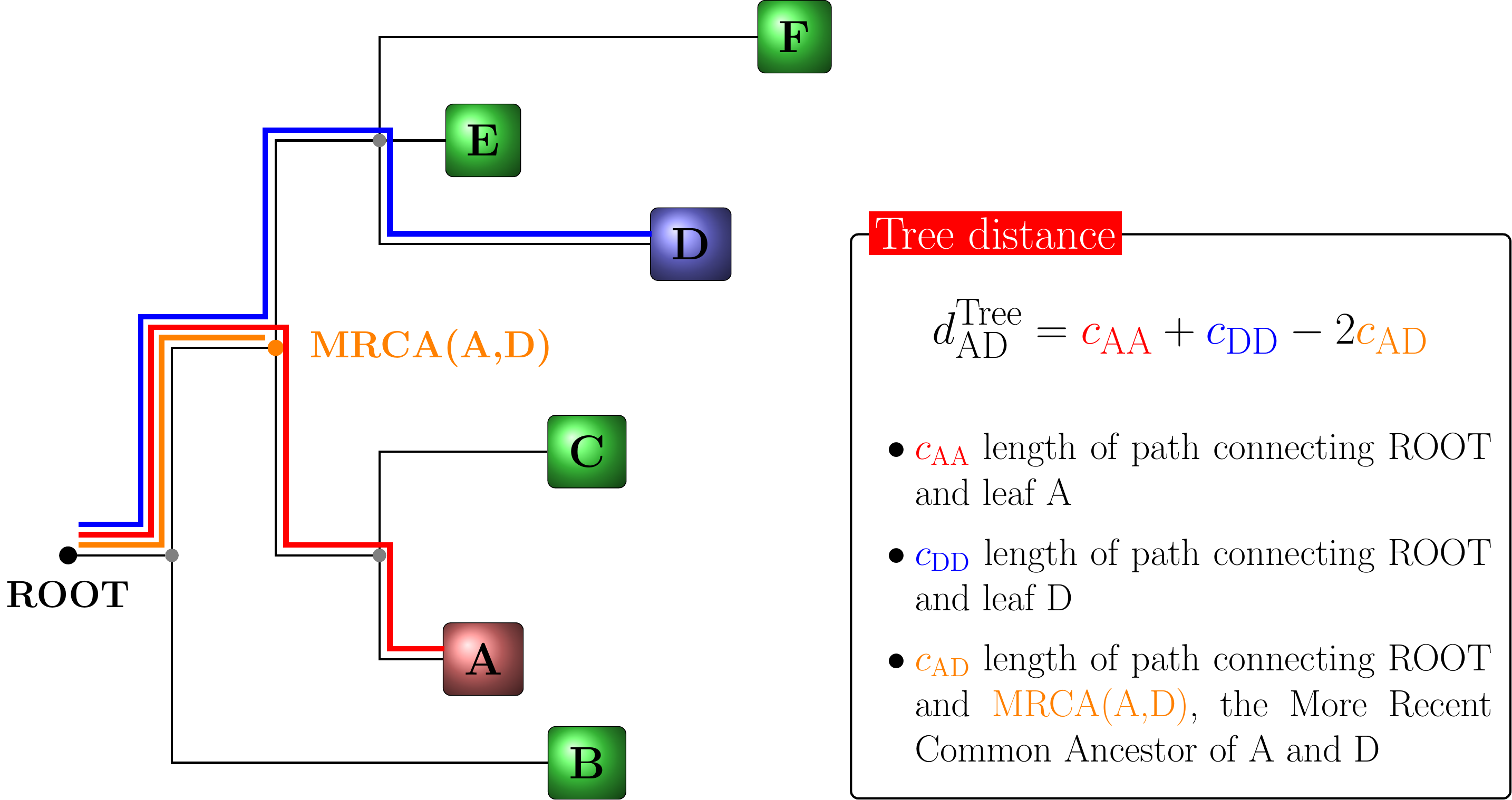} & 
\includegraphics[width=0.4\textwidth]{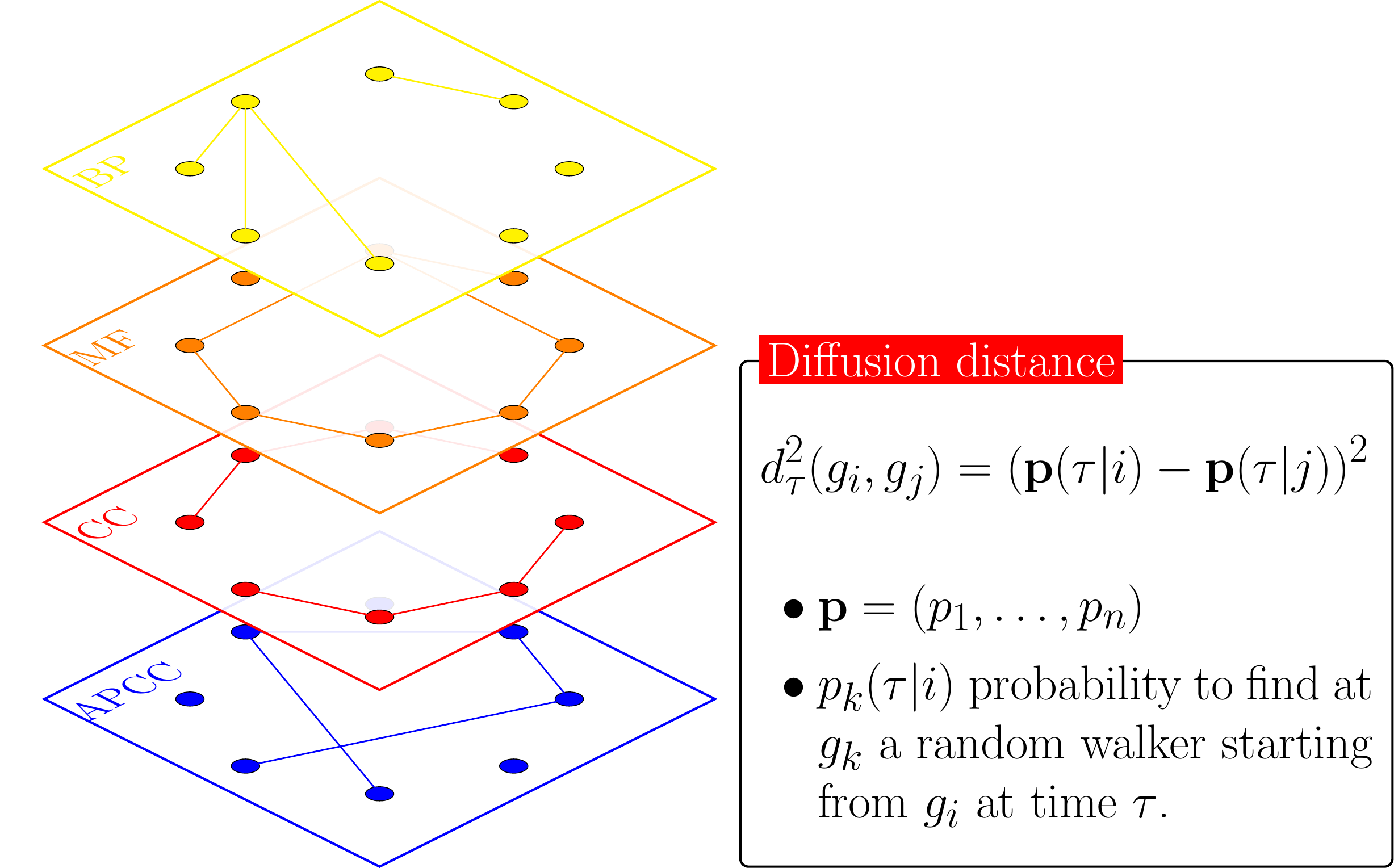}\\
(a) & (b)
\end{tabular}
\caption{(a) Patristic distance on a tree; (b) Diffusion distance for every layer of the multilayer networks for transcriptomics data: the four layers are Absolute Pearson Correlation Coefficient (APCC) and the three Gene Ontology layers Cellular Component (CC), Molecular Function (MF) and Biological Process (BP).}
\label{fig:metrics}
\end{figure}

Metagenomics data are usually described in terms of operational taxonomic units (OTUs), a general definition to identify groups of closely related individuals (e.g. species, genus, \ldots). OTUs have a well-defined structure and this provides our first example because a proximity measure can indeed be inherited by the tree structure connecting the OTUs. In general terms  any tree distance can be chosen as base metric for the convolution functions; here we demonstrate our approach on the patristic distance (see Fig.~\ref{fig:metrics}(a)), \textit{i.e.}, the sum of the lengths of all branches connecting two OTUs on the phylogenetic tree~\cite{stuessy08patrocladistic}.

Our second domain is transcriptomic data, \textit{i.e.} gene expression, for which several definitions of distances between two genes have been proposed. 
Most of such definitions are taking into account the genetic distance on the genome, or statistics of the gene expression. 
We need to add that the very fast development of biology of regulation~\cite{fantom14promoter} and about the implication of chromatin structure on gene expression~\cite{singh16deepchrome}, and most recent findings in measuring differences across cell types and functions~\cite{gtex17genetic} make gene expression metrization a moving target. 
In particular, experiments are currently running on both microarray and RNA-Seq data, with the MAQCII/SEQC Neuroblastoma dataset~\cite{zhang15comparison,oberthuer15revised} (498 samples on both platforms) as the first testbed.
Here we propose a principled metric for gene expression, relying on the mathematics of multilayer networks~\cite{kivela14multilayer}. 
The metrics can be applied to other omics data types, and extended to deal with integration. 
In details, a multilayer undirected weighted network on shared vertices (\textit{i.e.}, an edge-colored graph) is built with genes as the nodes. 
This graph includes 4 layers: in the first layer, two nodes are connected by an edge if the absolute Pearson correlation between the expression across all samples of the corresponding genes is larger than a significance threshold; in this case, the weight of the edge is given by the absolute value of the correlation coefficient. 
The other three layers are built according to the three categories of the Gene Ontology, i.e., cellular component, molecular function and biological process~\cite{ashburner00gene,gene17expansion}. 
In each layer, a link connects two genes if their semantic similarity (\textit{e.g}. as defined by GOSemSim~\cite{yu10gosemsim}) is not null, and the edge weight is exaclty this non-zero value. 
Given this multilayer structure, the metric between two genes on each layer is given by the diffusion distance~\cite{dedomenico17diffusion} (see Fig.~\ref{fig:metrics}(b)), which quantifies the likelihood that two random walkers originating in the two genes meet in some other node of the net. 
The overall distance in the network is given by the $L_2$ Euclidean product of the diffusion distances of each layer.
 
Once a omics distance is defined, the concept of nearest neighbours among features comes straightforwardly and the CNN algorithm can be applied.
However, by definition, the output of a CNN consists of linear combinations of the original input features: this implies that, after the first CNN layer, the problem of finding the neighbours of a feature becomes the hardest task of finding the neighbours of a linear combination of features.
We thus propose to map features into points of a $k$-dimensional metric space preserving distances as well as possible via a MultiDimensional Scaling (MDS) projection~\cite{cox01multidimensional}, which can be applied if the starting distance is Euclidean.
A further refinement is provided by sparsifying MDS via regularized low rank matrix approximation~\cite{shen07sparse} through the addition of the smoothly clipped absolute deviation penalty~\cite{fan01variable}, tuned by cross-validation.
 
\begin{figure}[!t]
\centering\includegraphics[width=0.9\textwidth]{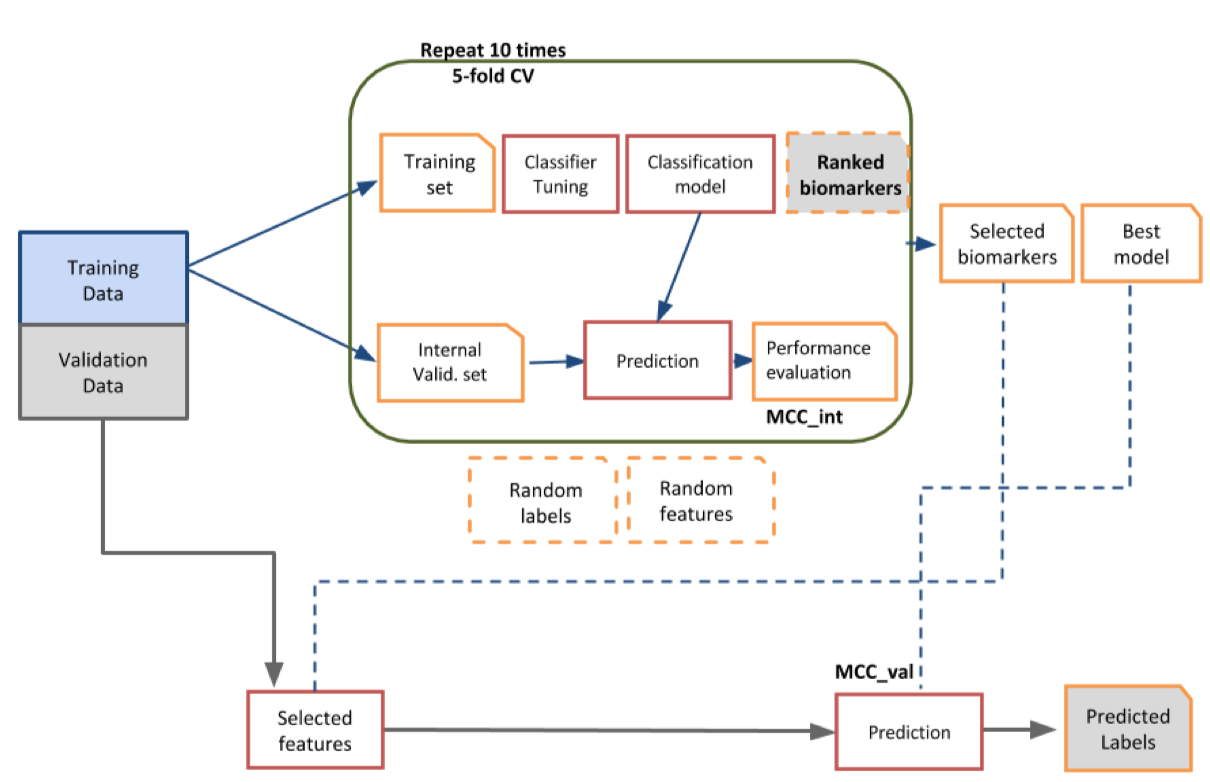}
\caption{{Data Analysis Protocol for the experimental framework.}}
\label{fig:dap}
\end{figure}
 
\section{OmicsCNN and experimental setup}
\label{sec:omicscnn}
The convolutional layer combined with the neighbours detection algorithm is operatively implemented as a novel Keras layer~\cite{chollet15keras} called OmicsConv, which convolves the selected neighbours with the filters.
In our metagenomics case study, the number of neighbours and the number of filters are both set to 16.
In a generic OmicsConv layer, the structure input is represented by a collection of metafeatures, \textit{i.e.} linear combinations of the original features; the first OmicsConv layer has the original set of features as input.
The OmicsConv is then repeated; finally, the terminating layers of the OmicsCNN are a MaxPooling, then a Flatten layer and, finally, a Fully Connected with 128 neurons and a 0.25 Dropout.
Each convolutional layer has a Scaled Exponential Linear Units (SELU) \cite{klambauer17selfnormalizing} as the activation fuction,and the dense layer in transfer learning experiments uses a sigmoid activation function.
Adam~\cite{kingma14adam} is used as optimizer with learning rate 0.0005.
The global learning architecure is called OmicsCNN, shown in Fig.~\ref{fig:omicscnn}.
 
To ensure predictive power and limit overfitting effect, the experimental framework is structured following the guidelines recommended by the US-FDA led studies MAQC/SEQC~\cite{maqc10maqc,seqc14comprehensive} that investigated the development of predictive models for the analysis of high-throughput data.
The experimental setup is realized as a 10$\times$5-fold cross-validation schema with a feature selection and ranking procedure, implementing a Data Analysis Protocol (DAP) to control for selection bias and other overfitting effects and warranting honest performance estimates on external validation data subsets. 
Top ranking features are recursively selected as the $k$-best at each round, and finally aggregated via Borda algorithm~\cite{jurman12algebraic}.
Model performance is computed for increasing number of best ranking features by Matthews Correlation Coefficient (MCC), the measure that better convey in an unique value the confusion matrix of a classification task, even in the multiclass case~\cite{matthews75comparison,baldi00assessing,gorodkin04comparing,jurman12comparison}.
Experiments with randomized features and labels are also performed as model sanity check.
 
The OmicsCNN is implemented as a Keras v2.0.8 layer, while the whole DAP is written in Python/Scikit-Learn.
All computations were run on a Microsoft Azure platform with 2x {NVIDIA} Tesla K80 GPUs.
A first version of the architecture, dedicated to metagenomics application, together with the IBD and synthetic metagenomics datasets are available at \url{https://gitlab.fbk.eu/MPBA/phylogenetic-cnn}.
 
\section{Case study}
\label{sec:case}
Sokol's lab data~\cite{sokol17fungal} collects microbiome information with 16S sequencing for 38 healthy subjects (HS) and 222 inflammatory bowel disease (IBD) patients.
The bacterial composition was analysed using 16S sequencing and a total number of 306 different OTUs was found.
IBD is a complex disease arising as a result of the interaction of environmental and genetic factors inducing immunological responses and inflammation in the intestine and primarily including ulcerative colitis (UC) and Crohn’s disease (CD).
Both disease classes are characterized by two conditions: flare (f), when symptoms reappear or worsen, and remission (r), when symptoms are reduced or disappear.
Finally, since CD can affect different parts of the intestine, we distinguish ileal Crohn’s disease (iCD) and colon Crohn’s disease (CD).
Note however,that the number of non zero features varies for the different rom tasks to task, (defined by disease, condition site) since some features may vanish on all samples of a class.
Due to the paucity of samples preventing an effective deep learning training phase, a synthetic dataset $\textbf{D}$ was generated to be used as the training set for the domain adaptation task.
Operatively, $\textbf{D}$ is generated as compositional data whose structure resembles the IBD data, sampling 10000 points from multivariate normal distributions with given covariances and mean.

OmicsCNN is first trained on the synthetic dataset $\textbf{D}$, then all layers but the last one are freezed, the last layer is substituted by a 2-neurons Dense layer and then retrained on the IBD dataset.
The full DAP is then applied to the IBD datasets, comparing the performance with standard (and shallow) learning algorithms such as linear Support Vector Machines (SVM) and Random Forest (RF), and with a standard Multi Layer Perceptron (MLPNN)~\cite{bishop95neural}, obtaining the results listed in Tab.~\ref{tab:ext}. 

\begin{table}[!t]
\caption{Dataset $\textbf{D}$ on IBD: classification performances of OmicsCNN compared to other classifiers on the external validation dataset.}
\label{tab:ext}
\begin{center}
\begin{tabular}{c|cccc}
Task & OmicsCNN & LSVM & MLPNN & RF \\
\hline
UCf & \textbf{0.741} & 0.740 & 0.666 & 0.699\\
UCr & 0.583 & 0.497 & 0.608 & \textbf{0.678}\\
CDf & \textbf{0.858} & 0.642 & 0.705 & 0.707 \\
CDr & \textbf{0.853} & 0.654 & 0.654 & 0.597 \\
iCDf& 0.842 & 0.418 & 0.401 & \textbf{0.920} \\
iCDr& \textbf{0.628} & 0.414 & 0.414 & 0.418 \\
\end{tabular}
\end{center}
\end{table}
OmicsCNN achieves better results in 4 tasks out of 6, with interesting improvements in the hardest CD/iCD tasks, suggesting even better outcomes are to be expected with more intensive optimization of the components of the proposed OmicsCNN architecture.
 
\section{Conclusions} 
\label{sec:conclusions}
We introduced here the first steps towards a global plan of making CNNs available to a number of (omics) data types by defining theoretically grounded embedding of features into metric spaces to activate a nearest neighbour function.
Encouraging initial results on metagenomics and transcriptomics support further investigations and extension to other metrizable omics structures.

\bibliographystyle{unsrtnat}
\bibliography{jurman17omics}
\end{document}